\documentclass[twocolumn,showpacs,preprintnumbers,amsmath,amssymb,pra]{revtex4}
\usepackage{graphicx}% Include figure files
\usepackage{dcolumn}% Align table columns on decimal point
\usepackage{bm}% bold math
\usepackage{natbib}
\begin{document}
%\begin{large}
\title{Comment on ``Prospect of optical frequency standard based on a $^{43}$Ca$^+$ ion"}

\author{C. Champenois}\email{Caroline.Champenois@up.univ-mrs.fr}
\author{M. Knoop}
\author{M. Houssin}
\author{G. Hagel}
\author{M. Vedel}
\author{F. Vedel}

\affiliation{Physique des Interactions Ioniques et Mol\'eculaires
(CNRS UMR 6633), Universit\'e de Provence, Centre de Saint
J\'er\^ome, Case C21, 13397 Marseille Cedex 20, France}%
 \homepage{http://www.up.univ-mrs.fr/ciml/}

\date{\today}

\begin{abstract}
A recent evaluation of the frequency uncertainty expected for an
optical frequency standard based on a single trapped $^{43}$Ca$^+$
ion was published in Phys. Rev. A {\bf 72} (2005) 043404. The
paper contains some interesting information like systematic
frequency shifts but fails to depict their uncertainty, leading to
confuse accuracy and precision. The  conclusions
 about the major contribution to the frequency shift
are not consistent with the presented calculations and omit
comparisons with data published previously.
\end{abstract}


\maketitle

Optical frequency standards based on single trapped ions may reach
ultimate performances which are orders of magnitude better than
the  existing cesium clocks in the microwave domain
\cite{riehle04}. A recent evaluation of residual effects in a
frequency standard based on a single $^{43}$Ca$^+$ ion  predicts
an ultimate attainable precision below 10$^{-15}$ \cite{kajita05}.
This evaluation  takes into account shifts due to local fields
neglecting their uncertainty.  In this comment we  show that
a more complete and precise analysis of the residual field effects
is necessary to predict the potential frequency uncertainty
\cite{champenois04}.

\begin{figure}
\includegraphics[width=90mm]{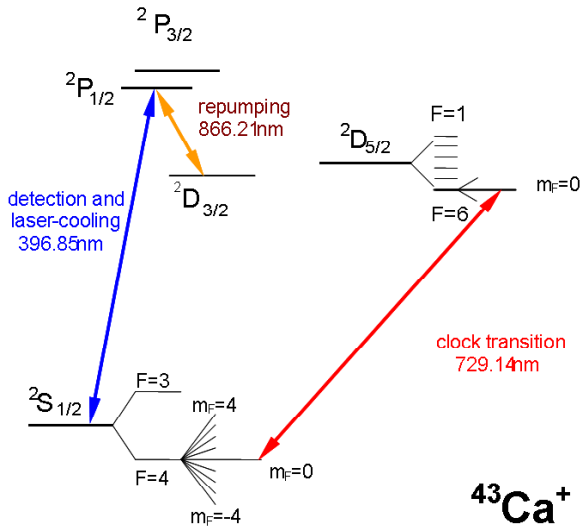}
\caption{Lower energy levels of the $^{43}$Ca$^+$ ion. All
wavelengths required for laser-cooling and interrogation of the
ion can be generated by solid-state lasers.}\label{fig:levscheme}
\end{figure}

The proposal of a single  $^{43}$Ca$^+$ ion stored in an rf trap
as an optical  frequency standard is based on its ultra-narrow
($\Delta\nu_{nat} < 160$ mHz) electrical quadrupole clock
transition at 4.1$\times 10^{14}$ Hz and its relatively simple
energy level scheme with all wavelengths in the optical domain,
allowing the concept of an all solid-state laser set-up (cf.
figure \ref{fig:levscheme}).

The  advantage of using the very rare (0.135\% of natural
abundance) odd isotope $^{43}$Ca$^+$ compared to the most abundant
$^{40}$Ca$^+$ relies in the possibility of eliminating the first
order Zeeman shift by using a $m=0$ sublevel, while the second order
Zeeman shift is minimized by the choice for the clock transition
of the hyperfine levels
$\left|4S_{1/2}, F=4 \right>$ and $ \left|3D_{5/2}, F=6 \right> $
 \cite{champenois04,kajita05}.
Although the value of the  magnetic field and its fluctuations
remain an important issue as this may be one of the major causes
of uncertainty in the frequency shifts,  the evaluation of this
uncertainty is not  addressed in \cite{kajita05}. A minimum
magnetic field is indeed required to split the chosen transition
from the two closest ones $\left|4S_{1/2}, 4, \pm1\right>
\rightarrow \left|3D_{5/2}, 6, \pm1 \right> $, however the choice
made in \cite{kajita05} of a 0.2~$\mu T$ (2 mG) magnetic field is
not motivated and the expected fluctuations of such a field are
not given. Magnetic field fluctuations as big as 0.2~$\mu T$ over
one day have been observed in an unshielded environnement
\cite{bize03}. Then, exploiting the linear Zeeman shift of the
$\left|4S_{1/2}, 4,0 \right> \rightarrow \left|3D_{5/2}, 6, \pm 2
\right> $ transitions like suggested in \cite{kajita05} may not be
sufficient to correct these fluctuations over one day. As the
Zeeman effect is quadratic, its fluctuations depend linearly on
the strength of the magnetic field and on the amplitude of its
fluctuations. With a 0.2~$\mu T$ magnetic field and 0.2~$\mu T$
fluctuations, the Zeeman frequency shift uncertainty is twice the
Zeeman shift itself and reaches 0.72~Hz. In  fact, the choice of
the magnetic field results from a compromise between maintaining a
high level of fluorescence \cite{gill03}, splitting the sublevel
transitions and keeping the Zeeman effect fluctuations low. A
complete description of the magnetic field (average value and
fluctuations) is therefore needed to estimate the uncertainty
induced by the Zeeman effect, which is missing in \cite{kajita05}.

The other significant effect causing frequency shift is the Stark
effect. It can be evaluated through the polarizability of the
states involved in the clock transition. The calculation presented
in \cite{kajita05} is however only partial. On the one hand,  the
anisotropic contribution (tensorial part) of the $D_{5/2}$
polarizability is not mentioned. This contribution is not relevant
as long as the major contribution to the local electric field is
caused by the black-body radiation, as considered in
\cite{kajita05}. Nevertheless,  when one considers cooling the
vessel to reduce this contribution, as realized in the
$^{199}$Hg$^{+}$ frequency standard \cite{rafac00}, this
anisotropic contribution must be taken into account
\cite{champenois04}. On the other hand, reference \cite{kajita05}
gives no estimation of the uncertainty of the calculated
polarizability. The polarizability of the $S_{1/2}$ state is well
known since the sum of the oscillator strengths taken into account
is very close to 1.  On the contrary, as pointed out in
\cite{champenois04}, the   $D_{5/2}$ state polarizability has a
large error bar because the sum of all the oscillator strengths of
the known transitions is only 0.48. The uncertainty of the
polarizability is therefore almost as big as the estimated
polarizability itself and it induces a non-negligible uncertainty
of the Stark shift. We found that, for a vessel at 300 K, the
Stark shift is 0.39 Hz, which agrees with the result of
\cite{kajita05}, but we showed that the uncertainty of this shift
can be as big as $\pm 0.27$ Hz, and can become the major
contribution in the overall error budget. As for the effect of the
coupling of the electric quadrupole moment of the $3D_{5/2}$ state
to any electric field gradient, its contribution to the
uncertainty can be reduced to the 0.1 Hz level
\cite{kajita05,champenois04} by measuring the transition frequency
for three orthogonal directions of the magnetic field, like first
proposed in \cite{itano00}.

In a thorough investigation of line-broadening effects, the AC
Stark (or light) shift depending on the laser intensity used to
probe the clock transition at 729 nm should also be evaluated.
Actually, for a Rabi frequency of 1000 $s^{-1}$ and a detuning
smaller than $\pm 10$ Hz to probe the low- and the high-frequency
side of the clock transition, the effect may be as small as $\pm$6
mHz for a magnetic field of 0.1 $\mu$T \cite{champenois04}.

In summary,  the overall uncertainty budget  given in
\cite{kajita05} is not complete. First, the most significant
frequency shift does not come from the quadratic Zeeman effect
since the static Stark effect is as big as this last effect (0.40
Hz compared to -0.36 Hz). Second,  the uncertainty of the
frequency shifts must be evaluated to estimate the precision of
the clock \cite{riehle04}, which is  missing in \cite{kajita05}.
In our previous paper \cite{champenois04}, we conclude that, if at
300 K, the major source of relative frequency uncertainty ($\pm 9
\times 10^{-16}$) would be due to the Stark effect,  for a vessel
cooled down to 77 K, the uncertainty  would result from the
fluctuations of the quadratic Zeeman effect and from the
experimental uncertainty in pointing three orthogonal directions
for compensating the quadrupole shift. It was estimated to $\pm 4
\times 10^{-16}$ (relative uncertainty) with room for improvement.
All the contributions to the systematic frequency shift and its
uncertainty are summarized in table \ref{tab_precision} which is
reproduced from reference \cite{champenois04}.

\begin{widetext}
\begin{table}[bht]
\caption{Uncertainty budget for the frequency transition of
$\left|S_{1/2},\ 4, \ 0\right> \rightarrow \left|D_{5/2},\ 6,\
0\right>$ in $^{43}$Ca$^{+}$ \cite{champenois04}}
\label{tab_precision} \hspace{-2.cm}\begin{tabular}{|c|c|c|c|}
\hline
effect & fields/conditions &  shift (Hz)@ 300 K & @ 77 K  \\
\hline \hline
second order Zeeman effect & 0.1 $\mu$T & $-0.09 \pm 0.09$ & $-0.09 \pm 0.09$ \\
Stark effect & radiated and bias static field   & $+0.39 \pm 0.27$  &  $\leq$ 0.012 \\
$D_{5/2}$ coupled to the field gradient & 1 V/mm$^2$ & $\pm 0.1$ & $\pm 0.1$\\

AC Stark effect @ 729 nm & 0.75 $\mu$W/mm$^2$, 0.1 $\mu$T & $\pm 0.006$& $\pm 0.006$\\
second order Doppler effect & ion cooled to the Doppler limit &
$-1 \times 10^{-4}$& $-1 \times 10^{-4}$ \\
\hline
global shift and uncertaintity  & & +0.3 $\pm 0.4$  & -0.09 $\pm$ 0.19\\
\hline
relative shift and uncertaintity & & $+7 (\pm 9)\times 10^{-16}$ & -2 ($\pm$ 4)$\times 10^{-16}$\\
\hline
\end{tabular}
\end{table}
\end{widetext}
%\end{large}

\end{document}